\documentclass[12pt,onecolumn,draftcls]{IEEEtran}

\usepackage{amsmath}
\usepackage{cite}
\usepackage[margin=26mm]{geometry}
\usepackage[final]{graphicx}
\usepackage{bm}
\usepackage{amssymb}
\usepackage{color}

\usepackage{epsfig}
\usepackage{nicefrac}
\usepackage{algorithm}
\usepackage[noend]{algpseudocode}

\DeclareMathOperator*{\argmin}{arg\,min}

\title{Non-Linear Relay Optimization using Deep-Learning tools}
\author{Itsik Bergel\thanks{\noindent Itsik Bergel (itsik.bergel@biu.ac.il) is with the Faculty of Engineering, Bar Ilan University, Ramat-Gan, Israel.}}

\date{October 2022}

\begin{document}

\maketitle

\begin{abstract}
Widespread deployment of relays can yield a significant boost in the throughput of forthcoming wireless networks. However, the optimal operation of large relay networks is still infeasible. 
This paper presents two approaches for the optimization of large relay networks. In the traditional approach, we formulate and solve an optimization problem where the relays are considered linear. In the second approach, we take an entirely new direction and consider the true non-linear nature of the relays. Using the similarity to neural networks, we leverage deep-learning methodology. 

Unlike previous applications of neural networks in wireless communications, where neural networks are added to the network to perform computational tasks, our deep relay optimization treats the relay network itself as a neural network. By exploiting the non-linear transfer function exhibited by each relay, we achieve over $15$dB gain compared to traditional optimization methods. Moreover, we are able to implement part of the network functionality over the relay network. Our findings shed light on the potential of deep relay optimization, promising significant advancements in future wireless communication systems.
\end{abstract}

\section{Introduction}
The importance of relaying has been known since the early days of wireless communication (e.g., \cite{harmon1912girdling,cover1979capacity,laneman2004cooperative}). In recent years, relays have become even more accessible due to progress in energy harvesting and full duplex communication.

Energy harvesting (e.g., \cite{nasir2013relaying,lu2014wireless}) enables devices to gather energy from their surroundings without relying on a traditional power source. The limited energy output is generally sufficient to activate a low-power relay. Full duplex communication (e.g., \cite{riihonen2011mitigation,sabharwal2014band}) allows devices to transmit and receive signals over the same frequency simultaneously. Thus, current technology enables the deployment of many relays without worrying about power sources or wasting valuable spectral resources.

However, despite the potential for substantial throughput gains, current communication networks underutilize relays. This is primarily due to the complexity involved in managing and optimizing a large number of relays.

We commonly distinguish between two types: decode-and-forward relays (e.g., \cite{nabar2004fading, kramer2005cooperative, nazer2011compute}) and amplify-and-forward relays (e.g., \cite{nabar2004fading, zhao2006improving}).
Decode-and-forward relays can achieve superior performance, but at a cost of significantly higher relay complexity and more challenging network design.

In this research, we focus on amplify-and-forward relays, which boast simpler manufacturing and operation. Optimization of amplify-and-forward relay networks is far from trivial, as their performance usually presents a non-convex behavior. Recent works (e.g., \cite{good_phan2012beamforming,sanguinetti2012tutorial}) have made progress and solved the optimization of various relay network topologies. Yet, there is no known solution for general relay networks.

Our work addresses this problem from two distinct perspectives. The first part extends the state-of-the-art relay network optimization, by solving the optimization of cascade relay networks (encompassing also all topologies with known optimal solutions). This novel optimization approach offers a versatile solution applicable to any relay network without loops. This solution stands as an independent novel contribution, presenting a substantial extension beyond existing results. Moreover, it provides a benchmark for evaluating the novel scheme of the second part.

The second part of our work steps out of traditional network optimization and presents a completely novel approach to relay design.

One key limitation of traditional relay network optimization lies in considering the relays as linear amplifiers (with a power constraint). Thus, they must limit the relay operation to the regime where it can be approximated as linear. As a result, these methods are forced to set a power constraint lower than the actual power achievable by the relays.

In our research, we recognize and leverage the non-linear transfer function exhibited by each relay. Recognizing the striking similarity between relay networks and neural networks, we embrace neural network algorithms to effectively optimize the relay network. By doing so, we unleash the relay network's performance beyond the confines of linearity, enhancing communication throughput and efficiency.

Neural networks have gained much popularity in recent years due to their ability to solve tough computational challenges, particularly when a good problem model is unavailable. The use of neural networks has been suggested in many communication applications (e.g., \cite{chen2019artificial,nikbakht2020unsupervised,sholev2020neural,saxena2021reinforcement}) and even in relay applications (e.g., \cite{zhang2020neural,xu2021intelligent,guo2021energy}). In particular, \cite{zhang2020neural} used neural networks for relay selection, \cite{guo2021energy} combined them with power allocation, and \cite{xu2021intelligent} used neural networks for predicting outage probabilities.

However, all applications of neural networks in wireless communications focus on inserting neural networks into nodes in the network to perform computational tasks.

In this work, we use neural network technology in a completely different way. We observe that the power limit at the relay exhibits a non-linear transfer function. This non-linear behavior is very similar to the hyperbolic tangent (tanh) commonly used in neural networks. Thus, instead of adding a neural network to our system, we treat our system as a neural network. Thus we analyze and optimize the relays using neural network algorithms.


Furthermore, we extend the benefits of the similarity to neural networks by implementing various functionalities over the relay network. This opens up exciting possibilities for data processing over the relay network. We demonstrate these capabilities by training the network to non-linearly separate the transmitted signal into the desired components of each receiver.
Our results show gains of over 15 dB compared to the state of the art in a cellular network with 100 relays.

The main advantages of the \emph{deep relay (DR) optimization} are:
\begin{itemize}
\item Robust optimization, using well-established deep-learning techniques.
\item Improved communication over relay networks.
\item New computational capabilities "over the air".
\end{itemize}

The main differences between the DR and other implementations of neural networks are:
\begin{itemize}
\item No added neurons - Relays are treated as neurons but are actual parts of the network.
\item Limited control over the network topology - Most of the network topology is determined by the channel gains, resulting from physical phenomena. The network optimization can only control the gain (and possibly bias) at each relay.
\item Noisy "neurons" - The input of each relay is affected by additive noise. Thus, we need to cope with many noise sources within the network.
\end{itemize}

While the proposed approach is relevant for all applications of relay networks, we prefer to focus here on low-complexity receivers. Such receivers are important, for example, for Internet-of-Things (IoT) devices \cite{akpakwu2017survey,herrero2022fundamentals,souri2022systematic}, where each receiver requires a low data rate, but the network must support a large number of receivers.

It is important to note that the relay network can learn several scenarios simultaneously. Using just a low rate control signaling, the network can switch between a set of scenarios, where each relay keeps in memory its gain and bias for each scenario. Hence, the same network
 can serve a large number of users, keeping a specific setup in memory for each set of users.

In conclusion, this research delves into the untapped potential of non-linear relay networks, leveraging deep-learning tools to optimize their performance. By treating the relay system as a neural network, we unlock greater power utilization, leading to improved communication efficiency.

In the following, we first present the system model in Section II. Section III solves the optimization problem in the traditional approach, while Section IV presents our novel deep-learning approach. Section V presents numerical studies that demonstrate the advantages of our approach and Section VI gives our concluding remarks.

\section{System model}
\begin{figure}[t]
\center\epsfig{figure=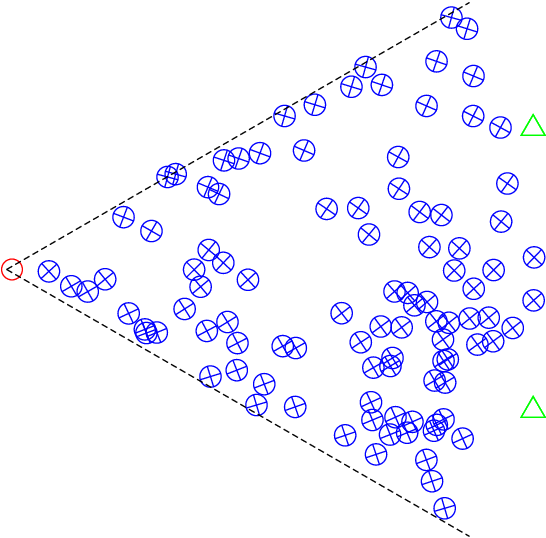,width=0.45\textwidth,trim=0 0 0cm 0, clip}
 \caption{\label{fig:prelim_system_model} Network model example: One BS (red circle), 100 relays and 2 receivers (green triangle). As shown in the figure, the receive antennas of all relays are pointing toward the BS.}
\end{figure}

We consider a single sector of a cellular network, with a single transmitter (base station), $M$ receivers and $N$ relays as demonstrated in Fig. \ref{fig:prelim_system_model}. All transmissions in the network are performed over the same frequency. The transmitter simultaneously transmits independent data to each of the receivers. 

For the clarity of this basic study on deep relay networks, we take two simplifying assumptions. We assume that all signals are real (i.e., transmitted signals, channel gains and relay gains are all real) and we assume perfect full duplex and directional antennas. These assumptions simplify the mathematical presentation while retaining the essence of the relay network. Both assumptions should be relaxed in future studies.

We note that traditional optimization is performed by maximizing the signal to noise ratio (SNR). Thus, it does not depend on the specific modulation used. On the other hand, deep-learning is based on the specific signal values, and hence must focus on a specific modulation. In the following, we describe the complete signal structure, recalling that the definitions of the modulation
are only needed for Section \ref{sec:DR_optimization}.

The bits intended for receiver $m$ at time $k$ are denoted by $\mathbf{u}_m[k]=\big[u_{m,1}[k],\ldots,u_{m,B}[k]\big]$ where $B$ denotes the number of bits simultaneously transmitted to each receiver.
We consider a single antenna transmitter (and hence, signal separation cannot be done by spatial multiplexing). Thus, all transmitted data is jointly modulated into a single symbol. This is done by stacking the bits intended for all users into a single vector, $\mathbf{u}[k]=[\mathbf{u}_1[k],\ldots,\mathbf{u}_M[k]]$, using gray code and then pulse amplitude modulation (PAM) modulation. Without loss of generality, the maximal absolute value at the transmitter output is set to $1$.

More specifically, let $a=0,\ldots,2^{MB}-1$ represent the symbol index, then the transmitted value for symbol $a$ is $(2a-2^{MB}+1)/(2^{MB}-1)$. This value represents a value of $\left\lfloor \frac{a+2^{c-1}}{2^{c}}\right\rfloor\mod 2$ for the $c$ transmitted bit ($1\le c\le MB$), which is bit $(B-(c-1 \mod B))$ of user $(M+1-\lceil c/B\rceil)$.

For example, for $B=1$ and $M=2$, the $4$ PAM points to deliver one bit per channel use for each user are shown in table \ref{table:modulation}.

\begin{table}[h]
\begin{center}
\begin{tabular}{ |c|c|c|c|c| } 
 \hline
 Bit for user 1, $u_{1,1}$ & 0 & 0 & 1 &1 \\ 
 Bit for user 2, $u_{2,1}$ & 0 & 1 &1 &0\\ 
 \hline 
 Transmitted value & -1 & -1/3&1/3&1 \\ 
 \hline
\end{tabular}
\end{center}
\caption{Constellation points for transmitting 1 bit per user for 2 users.} \label{table:modulation}
\end{table}
\begin{figure}[t]
\center\epsfig{figure=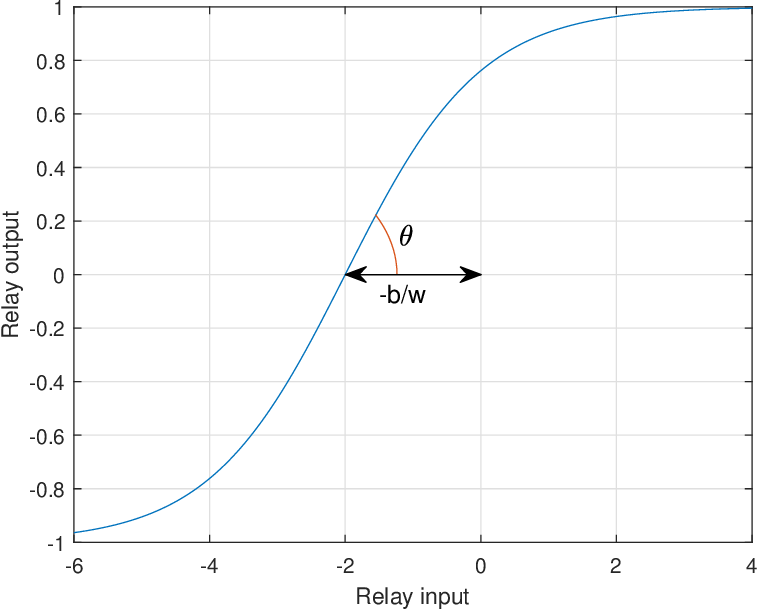,width=0.45\textwidth,trim=0 0 0cm 0, clip}
\caption{\label{fig:tr_func} Transfer function of a relay with a gain of $w$ and a bias of $b$. The slope at the linear regime is determined by the gain such that $w=\tan(\theta)$.}
\end{figure}

Each relay amplifies its received signal, applies its (non-linear) transfer function, and transmits it forward. The system is tuned by setting the gain of each relay, and (possibly) adding bias signals at the relays (to better utilize the relay non-linearity).

 Each relay is equipped with four directional (sector) antennas, each of $90^\circ$ width. Out of these antennas, only two are active. The receive antenna is pointing backward to the transmitter (and will receive also any relay in its beam width). The transmit antenna is pointing in the opposite direction (forward).

The assumption of ideal directional antennas guarantees 
that the network will not contain loops. Thus, we can look at it as a cascade network, in which the relays are divided into layers, and each relay can receive signals only from the BS and from relays at previous layers.
Let $y_{i,a}[k] $ be the signal received by relay $a$ of layer $i$ at time $[k]$, and $\mathbf y_i[k]=\big[y_{i,1}[k],\ldots,y_{i,N_i}[k]\big]^T$ be the vector of inputs for all relays at layer $i$ (with $N_i$ being the number of relays at layer $i$). The received signal vector at layer 1 is given by:
\begin{IEEEeqnarray}{rCl}\label{e:input_first_layer}
\mathbf{y_1}[k]=\mathbf{h_1} s[k]+\mathbf{n}_1[k]
\end{IEEEeqnarray}
where $ s$[k] is the signal transmitted by the BS, $\mathbf h_i$ is the vector of channel gains from the BS to the relays of layer $i$, and $\mathbf n_i[k]$ is the vector of additive white noise at each relay, which is assumed to have independent Gaussian distribution with zero mean and variance $\sigma^2$.

The amplify-and-forward relay has limited power. The traditional analysis considers the relays as linear amplifiers with a power constraint. To improve performance, we consider the actual transfer function of the relays. 

Each amplifier has its unique transfer function. Typically, the transfer function is almost linear for low input values and reaches a saturation for large input values. An example of a transfer function is depicted in Fig. \ref{fig:tr_func}. This transfer function has two controlled parameters: the gain, $w$, and an added bias, $b$. (The traditional analysis does not consider the bias as it cannot improve the performance in a linear model.) Again, without loss of generality, we set the maximal absolute value at the output of each transmitter to be $1$.

The resemblance of Fig \ref{fig:tr_func} to the common transfer function of a neuron in a neural network leads to the concept of using deep-learning tools for the optimization of the network. Thus, the network is tuned by setting the gain and bias of each relay, and we use backpropagation to optimize the network. This approach will be presented in Section \ref{sec:DR_optimization}. 

 The signal at relay $j$ of layer $i$ is amplified by a gain of $w_{j,k}$, added to a bias term $b_{j,k}$ and then subjected to the amplifier non-linearity. For simplicity, we assume throughout that all relays are characterized by the hyperbolic tangent function. Thus, the signal at the output of layer $i$ is given by:
\begin{IEEEeqnarray}{rCl}\label{e:relay_output}
\mathbf o_i[k] = \tanh\left(\mbox{diag}(\mathbf w_i)\mathbf y_i[k]+\mathbf b_i\right).
\end{IEEEeqnarray}
The input for layer $i>1$ is given by:
\begin{IEEEeqnarray}{rCl}\label{e:input_general_layer}
\mathbf y_i [k]=\mathbf h_i s[k] +\sum_{\ell=1}^{i-1}\mathbf F_{i,\ell}\mathbf o_\ell[k]+\mathbf n_i[k] 
\end{IEEEeqnarray}
where $\mathbf F_{i,\ell}$ is the matrix of channel gains from layer $\ell$ to layer $i$.
Finally, we assume no direct link from the BS to the receivers, and the received signal at user $m$ is:
\begin{IEEEeqnarray}{rCl}\label{e:XXX}
r_m[k]=\sum_{i=1}^d \mathbf g_{i,m}^T\mathbf o_i[k]+\tilde{ n}_m
[k].
\end{IEEEeqnarray}
where $d$ is the number of layers, $\mathbf g_{i,m}$ is the vector of channel gains from layer $i$ to receiver $m$ and $\tilde{ n}_m[k]$ is the additive Gaussian noise, again with variance $\sigma^2$.  

 We assume Gaussian fading over all channels, so that each channel gain is given by $r^{-\alpha} \cdot v$ where $r$ is the link length, $\alpha=4$ is the path loss exponent and $v$ is independent Gaussian fading with zero mean and unit variance.

The receiver applies a set of detection functions to produce bit estimates:
\begin{IEEEeqnarray}{rCl}\label{e:bit-estimates}
\hat u_{m,b}[k]=q_{m,b}(r_m[k]), \ m=1,..., M,\ b=1,...,B
\IEEEeqnarraynumspace
\end{IEEEeqnarray}
and the performance is measured by the bit error rate (BER) given by
\begin{IEEEeqnarray}{rCl}\label{e:BER1}
\epsilon_{m,b}=\Pr(\hat u_{m,b}\ne u_{m,b}).
\end{IEEEeqnarray}
We will focus on max-min BER optimization. Thus, our network performance metric will be 
\begin{IEEEeqnarray}{rCl}\label{e:XXX}
\epsilon=\max_{m,b} \epsilon_{m,b}.
\end{IEEEeqnarray}

If the relays were linear, each network output would have been a scaled and noised version of the transmitted signal. In such a case, we expect the receiver to employ a PAM receiver that matches the transmitted modulation. Note that the structure of the modulation is such that users' $m$ data can be decoded with $mB$-PAM receiver. Thus, only user $M$ truly needs an $MB$-PAM receiver.

 With non-linear relays, we can train the relay to perform some of the signal separation and hence allow simpler receivers. In particular, we may wish that all users will employ only $B$-PAM receivers.

While the implementation of PAM receivers is standard, we give here an exact description of our implementation. This description serves to better define the types of receivers we consider, and also as a preparation for the deep-learning scheme, which will use some of these functions. To accommodate all types of considered receivers, we divide the detection functions into three parts: scaling, output processing and decision. 

The scaling part linearly adjusts the received signal to match the receiver decision zones. Thus, the scaled output is given by:
\begin{IEEEeqnarray}{rCl}\label{e:XXX}
\bar r_m[k] = \bar w_m r_m[k] +\bar b_m.
\end{IEEEeqnarray}
The values of $\bar w_m$ and $\bar b_m$ for each output are determined by the optimization algorithm.

The output processing part takes the network output and extracts the bits of the specific user. 
Let $f(x)=2\sqrt{x^2+\epsilon^2}-1$, with $\epsilon=0.01$. Also, let $f^{(0)}(x)=x$ and $f^{(z)}(x)=f\big(f^{(z-1)}(x)\big).$ We consider two types of receivers. For standard receivers, user $m$ uses an $mB$-PAM receiver, and the output processing for its $b$-th bit is 
\begin{IEEEeqnarray}{rCl}\label{e:q_eval_per_bit}
\bar q_{m,b}[k]=f^{(mB-b)}\left(-\frac{2^{MB}-1}{2^{MB}}\cdot \bar r_{m}[k]\right).
\end{IEEEeqnarray}
 For low-complexity receivers, all users use $B$-PAM receivers, and the output processing for their $b$-th bit is $\bar q_{m,b}[k]=f^{(B-b)}\left(-\frac{2^{MB}-1}{2^{MB}}\cdot \bar r_{m}[k]\right)$.

The decision function in all cases is $\hat u_{m,b}[k]=1$ if $\bar q_{m,b}[k]<0$ and $0$ otherwise. Note in particular that in the specific case of low-complexity receivers and $B=1$ the decision functions for all receivers simplify to the scaled binary-phase-shift-keying (BPSK) receivers, that is: $\hat u_{m,1}[k]=1$ if $- \bar w_m r_m [k]- \bar b_m<0$ and $0$ otherwise.


\section{Traditional (linear model) optimization}\label{Sec:Linear}

The traditional approach treats the relays as linear amplifiers. To that end, we constrain the relay output power to a low enough level, such that the hyperbolic function can be reasonably approximated as linear. 
In mathematical terms, the linear model is obtained by replacing \eqref{e:relay_output} with 
\begin{IEEEeqnarray}{rCl}\label{e:linear_relay}
\mathbf o_i [k]\approx \mbox{diag}(\mathbf w_i)\mathbf y_i[k]+\mathbf b_i
\end{IEEEeqnarray}
and adding a constraint $E\big[o_{i,k}^2[k]\big]\le P_{\max}$.

Using \eqref{e:linear_relay} instead of \eqref{e:relay_output}, the complete network is linear. Thus, each receiver will receive a scaled version of the transmitted signal plus additive Gaussian noise. In such a scenario, BER minimization is obtained by weighted SNR maximization subject to the power constraint. In the linear model, the bias term consumes output power, but has no benefit for the network. Thus, we set $\mathbf{b}_i=0$ for all $i$. 

We need to solve the optimization problem:
\begin{IEEEeqnarray}{rCl}\label{e:linear_opt}
\max_{\{\mathbf{w}_i\}} \min_m & &\ \zeta_m\cdot\mbox{SNR}_m
\\\notag
\text{Subject }&&\text{to:}
\\\notag && E[o_{i,j}^2]\le P_{\max} \quad i=1,\ldots,d,\ j=1,\ldots,N_i
\end{IEEEeqnarray}
where 
\begin{IEEEeqnarray}{rCl}\label{XXX}
\mbox{SNR}_m=\frac{E\left[\left|[E[r_m[k]|\mathbf{u}_m[k]]\right|^2\right]}{\mbox{Var}(r_m[k]|\mathbf{u}_m[k])}.
\end{IEEEeqnarray}
 The SNR weights, $\zeta_m$ are chosen to balance the BER of the different users. 

The optimization problem in \eqref{e:linear_opt} is not convex, and its solution was not derived so far. The closest solution is the one derived by Phan et al. \cite{good_phan2012beamforming}. In the following, we extend this solution to the problem at hand. This extension includes: i) extending the solution to multiple relay layers by alternating minimization over the layers ii) extending the solution to a cascade network and iii) changing the optimization variables to allow a solution.

We start by rephrasing the problem as a power minimization problem and constructing the alternating minimization. Let $\underline{\mathbf w}_v=\{{
\mathbf w}_1,\ldots,{
\mathbf w}_{v-1},{
\mathbf w}_{v+1},\ldots,{
\mathbf w}_d\}$, we define the $v$-th layer min max power for a weighted SNR, $\eta$ as: 
\begin{IEEEeqnarray}{rCl}\label{e:min_power}
P_v(\underline{\mathbf w}_v,\eta)=&&\ \ \min_{{
\mathbf w}_v} \ \ \max_j E\big[o_{i,j}^2[k]\big]\\ \notag
$Subject to:$ && \quad 
\zeta_m\cdot\mbox{SNR}_m\ge \eta 
,\quad 1\le m\le M.
\end{IEEEeqnarray} 
Once we will solve \eqref{e:min_power}, the solution to \eqref{e:linear_opt} can be easily evaluated using Algorithm \ref{alg:1}.
Thus, the main challenge in solving \eqref{e:linear_opt} is to solve \eqref{e:min_power}. 
\begin{algorithm}[t]
\caption{Alternating minimization for minimal BER}
\label{alg:1}
\begin{algorithmic}[1]
\State Start with a feasible solution\Repeat: Loop over $1\le v\le d$ 
\State Find the maximal $\eta$ such that $P_v(\underline{\mathbf w}_v,\eta)\le P_{\max}$ and update $\mathbf w_v$ accordingly.
\Until Convergence
\State If the users' BERs are not equal, adjust the weights $\{\zeta_m\}$ and repeat from 1. 
\end{algorithmic}
\end{algorithm}

For that purpose, 
we construct a more efficient network description. Let the extended input of the $i$-th layer be $\tilde{\mathbf o}_i [k]= [\tilde{\mathbf o}_{i-1}^T[k],\mathbf o_i^T]^T[k]$ with $\tilde{\mathbf o}_0[k]=s[k]$. We also define $\tilde N_i=\sum_{u=1}^i N_i$, $\tilde{\mathbf g}_m=[0,\mathbf g_{1,m}^T,\ldots,\mathbf g_{d,m}^T]^T$, $
\tilde{\mathbf w}_i=[
\mathbf 1_{1+\tilde N_{i-1}}^T, \
\mathbf w_i^T]^T$, $\tilde{\mathbf C}_i=[\mathbf 0 ,\ \mathbf I_{N_i}]^T$, $\mathbf F_i=[\mathbf F_{i,1},\ldots,\mathbf F_{i,i-1}]$, and
\begin{IEEEeqnarray}{rCl}\label{e:XXX}
\tilde{\mathbf F}_i=\begin{bmatrix}1&\mathbf 0 
\\ \mathbf 0 &\mathbf I_{\tilde N_{i-1}} \\
\mathbf h_i&\mathbf F_i\\\end{bmatrix} 
\end{IEEEeqnarray}
To clarify, note that $\tilde{\mathbf F}_1=[1, \mathbf h_i^T]^T$. 

Using this notation, the extended signal at the input level $i$ is:
\begin{IEEEeqnarray}{rCl}\label{e:XXX}
\notag\tilde{\mathbf o}_i &=&
\mbox{diag}(\tilde{\mathbf w}_i)\tilde{\mathbf F}_i\tilde{\mathbf o}_{i-1}+\tilde{\mathbf C}_i \mathbf n_i
\\&=& \left(\prod_{i'=1}^{i} \mbox{diag}\left(\tilde{
\mathbf w}_{i'} 
\right) \tilde{\mathbf{F}}_{i'} \right)s
\notag\\&&+\sum_{u=1}^i \left(\prod_{i'=u+1}^{i} \mbox{diag}\left(\tilde{
\mathbf w}_{i'} 
\right) \tilde{\mathbf{F}}_{i'} \right)\mbox{diag}\left(\tilde{
\mathbf w}_u 
\right)\tilde{\mathbf C}_u \mathbf n_u
\notag
\\&=&\mathbf G_{1,i} s
+\sum_{u=1}^i \mathbf G_{u+1,i}\mbox{diag}\left(\tilde{
\mathbf w}_u 
\right)\tilde{\mathbf C}_u \mathbf n_u
\end{IEEEeqnarray}
where $\mathbf G_{u,i}=\prod_{i'=u}^{i} \mbox{diag}\left(\tilde{
\mathbf w}_{i'} 
\right) \tilde{\mathbf{F}}_{i'}$. The received signal at the $m$-th mobile is:
\begin{IEEEeqnarray}{rCl}\label{e:XXX}
r_m=\tilde{\mathbf g}_{m}^T\tilde{\mathbf o}_d+\tilde{ n}_m
.
\end{IEEEeqnarray}

The signal to noise ratio (SNR) at receiver $m$ is
\begin{IEEEeqnarray}{rCl}\label{e:SNR}
\mbox{SNR}_m&=&\frac{\left| \tilde{\mathbf g}_{m}^T \mathbf G_{1,d} \right|^2\sigma_s^2/\sigma^2
}
{1+\sum_{u=1}^d \left\|\tilde{\mathbf g}_{m}^T\mathbf G_{u+1,d}\mbox{diag}\left(\tilde{
\mathbf w}_u 
\right)\tilde{\mathbf C}_u\right\|^2}  .
\IEEEeqnarraynumspace\end{IEEEeqnarray}
where $\sigma_s^2=E[s^2]$. Defining also $\mathbf{e}_j$ to be an indicator vector for the $j$-th element, the output power of relay $j$ is: 
\begin{IEEEeqnarray}{rCl}\label{e:pk}
p_{j}&=&E[o_{i,j}^2]
=\left|\mathbf{e}_{j+1}^T \mathbf G_{1,d}\right|^2\sigma_s^2
\notag\\&&
+\sigma^2\sum_{u=1}^i \left\|\mathbf{e}_{j+1}^T \mathbf G_{u+1,d}\mbox{diag}\left(\tilde{
\mathbf w}_u 
\right)\tilde{\mathbf C}_u\right\|^2.
\end{IEEEeqnarray}

To allow 
the optimization with respect to the gains of layer $v$. We note that for $u<v<i$ we can write:
\begin{IEEEeqnarray}{rCl}\label{e:XXX}
\mathbf G_{u,i}=\mathbf G_{v+1,i} \mbox{diag}\left(\tilde{
\mathbf w}_{v} 
\right) \tilde{\mathbf{F}}_{v}\mathbf G_{u,v-1}.
\end{IEEEeqnarray}
This can be conveniently extended to $u\le v\le i$ by defining $\mathbf G_{u,i}=1$ for any $u>i$. Thus, the noise term at receiver $m$ divided by $\sigma^2$ (i.e., the denominator of \eqref{e:SNR}) can be written as:
\begin{IEEEeqnarray}{rCl}\label{e:XXX}
\Xi_m&=&1+\sum_{u=1}^{v-1} \Big\|\tilde{\mathbf g}_{m}^T\mathbf G_{v+1,d} \mbox{diag}\left(\tilde{
\mathbf w}_{v} 
\right) \tilde{\mathbf{F}}_{v}\mathbf G_{u+1,v-1}
\notag\\&&\cdot\mbox{diag}\left(\tilde{
\mathbf w}_u 
\right)\tilde{\mathbf C}_u\Big\|^2+ \left\|\tilde{\mathbf g}_{m}^T\mathbf G_{v+1,d}\mbox{diag}\left(\tilde{
\mathbf w}_v 
\right)\tilde{\mathbf C}_v\right\|^2  
\notag\\&&+\sum_{u=v+1}^d \left\|\tilde{\mathbf g}_{m}^T\mathbf G_{u+1,d}\mbox{diag}\left(\tilde{
\mathbf w}_u 
\right)\tilde{\mathbf C}_u\right\|^2  
\notag\\
&=&1+\sum_{u=1}^{v-1} \Big\|\tilde{
\mathbf w}_{v}^T \mbox{diag}\left(\mathbf G_{v+1,d}^T\tilde{\mathbf g}_{m} 
\right) \tilde{\mathbf{F}}_{v}\mathbf G_{u+1,v-1}
\notag\\&&\cdot\mbox{diag}\left(\tilde{
\mathbf w}_u 
\right)\tilde{\mathbf C}_u\Big\|^2+ \left\|\tilde{
\mathbf w}_{v}^T \mbox{diag}\left(\mathbf G_{v+1,d}^T\tilde{\mathbf g}_{m} 
\right)\tilde{\mathbf C}_v\right\|^2  
\notag\\&&+\sum_{u=v+1}^d \left\|\tilde{\mathbf g}_{m}^T\mathbf G_{u+1,d}\mbox{diag}\left(\tilde{
\mathbf w}_u 
\right)\tilde{\mathbf C}_u\right\|^2 
\notag\\
&=&1+\sum_{u=1}^{v-1} \left\|\tilde{
\mathbf w}_{v}^T \mathbf L_{v,u}\right\|^2+ \left\|\tilde{
\mathbf w}_{v}^T \mathbf L_{v,v} \right\|^2  
+\ell_v.
\end{IEEEeqnarray}
where $\ell_v\left({\mathbf g}\right)=\sum_{u=v+1}^d \left\|\mathbf g^T\mathbf G_{u+1,d}\mbox{diag}\left(\tilde{
\mathbf w}_u 
\right)\tilde{\mathbf C}_u\right\|^2 $, $\mathbf L_{v,u}\left({\mathbf g}\right)=\mbox{diag}\left(\mathbf G_{v+1,d}^T \mathbf g
\right) \tilde{\mathbf{F}}_{v}\mathbf G_{u+1,v-1}\mbox{diag}\left(\tilde{
\mathbf w}_u 
\right)\tilde{\mathbf C}_u$ for $u>v$ and $\mathbf L_{v,v}\left({\mathbf g}\right)=\mbox{diag}\left(\mathbf G_{v+1,d}^T\mathbf g 
\right)\tilde{\mathbf C}_v$. Defining also $\tilde{\mathbf W}_v=\tilde{\mathbf w}_v\tilde{\mathbf w}_v^T$ and $\mathbf L_{v}\left({\mathbf g}\right)=\sum_{u=1}^{v} \mathbf L_{v,u} \left({\mathbf g}\right)\mathbf L_{v,u}^T\left({\mathbf g}\right)$ we can write:
\begin{IEEEeqnarray}{rCl}\label{e:XXX}
\Xi_m&=&1+\mbox{Tr}\left[\tilde{
\mathbf W}_{v}^T \mathbf L_{v}\left(\tilde{\mathbf g}_{m} 
\right)\right]
+\ell_v\left(\tilde{\mathbf g}_{m} 
\right).
\end{IEEEeqnarray}
Similarly, the signal term, divided by $\sigma^2$, can be written as $\mbox{Tr}\left[\tilde{
\mathbf W}_{v}^T \mathbf Q_{v}\left(\tilde{\mathbf g}_{m} 
\right)\right]$ where 
\begin{IEEEeqnarray}{rCl}\label{e:XXX}
\mathbf Q_{v}\left({\mathbf g}\right)&=&\frac{\sigma_s^2}{\sigma^2}\mbox{diag}\left(\mathbf G_{v+1,d}^T{\mathbf g} 
\right) \tilde{\mathbf{F}}_{v}\mathbf G_{1,v-1}\mathbf G_{1,v-1}^T\tilde{\mathbf{F}}_{v}^T
\notag\\&&\cdot\mbox{diag}\left(\mathbf G_{v+1,d}^T{\mathbf g}
\right).
\end{IEEEeqnarray}
Thus, the weighted SNR constraint of \eqref{e:min_power}
is written as 
\begin{IEEEeqnarray}{rCl}\label{e:XXX}
\mbox{Tr}\left[\tilde{
\mathbf W}_{v}^T \mathbf B_{v,m}\right]
\ge 1
+\ell_v\left(\tilde{\mathbf g}_{m} 
\right) 
\end{IEEEeqnarray}
where
\begin{IEEEeqnarray}{rCl}\label{e:XXX}
\mathbf B_{v,m}=\frac{\zeta_m\mathbf Q_{v}\left(\tilde{\mathbf g}_{m} 
\right)}{\eta}-\mathbf L_{v}\left(\tilde{\mathbf g}_{m} 
\right). 
\end{IEEEeqnarray}
Using the same terminology, we can write \eqref{e:pk} as:
\begin{IEEEeqnarray}{rCl}\label{e:XXX}
p_j=\sigma^2\cdot\mbox{Tr}\left[\tilde{
\mathbf W}_{v}^T \mathbf A_{v,j}\right]+\sigma^2\ell_v\left(\mathbf{e}_{j+1}\right)
\end{IEEEeqnarray}
where 
\begin{IEEEeqnarray}{rCl}\label{e:XXX}
\mathbf A_{v,j}= \mathbf Q_{v}\left(\mathbf{e}_{j+1}\right)+ \mathbf L_{v}\left(\mathbf{e}_{j+1}\right).
\end{IEEEeqnarray}

To further simplify the optimization, we perform the optimization with respect to $\tilde{
\mathbf W}_{v}$ instead of $\mathbf w_v$. Thus, we need to add the constraint that $\tilde{
\mathbf W}_{v}$ is rank $1$ and also $(\tilde{
\mathbf W}_{v})_{j,j}=1$ for any $1\le j \le 1+\tilde N_{v-1}$. Hence, 
\eqref{e:min_power} can be rewritten as
\begin{IEEEeqnarray}{rCl}\label{e:B_const}
&&P_v(\underline{\mathbf w}_v,\eta)=\sigma^2 \cdot \min_{\tilde{
\mathbf W}_v} \  \max_j \mbox{Tr}\left[\tilde{
\mathbf W}_{v}^T \mathbf A_{v,j}\right]+\ell_v\left(\mathbf{e}_{j+1}\right)
\notag\\ 
&&\quad$Subject to:$  
\\&&\quad\quad\mbox{Tr}\left[\tilde{
\mathbf W}_{v}^T \mathbf B_{v,m}\right]
\ge 1
+\ell_v\left(\tilde{\mathbf g}_{m} 
\right) 
,\quad 1\le m\le M
\notag\\ \notag 
&& \quad\quad\tilde{\mathbf W}_v\ge 0, 
 \mbox{rank}(\tilde{\mathbf W}_v)=1
, (\tilde{
\mathbf W}_{v})_{j,j}=1, 
\\\notag&&\hspace{4.5cm}\quad 1\le j \le 1+\tilde N_{v-1}. 
\end{IEEEeqnarray}
 
Following \cite{good_phan2012beamforming} 
we replace the rank $1$ constraint by $\mbox{Tr} [\tilde{\mathbf W}_v]-\lambda_{\max}(\tilde{\mathbf W}_v)\le 0$, and further simplify the problem by moving the constraint to the utility function, resulting with:
\begin{IEEEeqnarray}{rCl}\label{e:mu_const}
&&P_v(\underline{\mathbf w}_v,\eta)=\sigma^2\cdot\min_{\tilde{
\mathbf W}_v} \ \ \max_j \mbox{Tr}\left[\tilde{
\mathbf W}_{v}^T \mathbf A_{v,j}\right]+\ell_v\left(\mathbf{e}_{j+1}\right)
\notag\\&&\hspace{2cm}+\mu \left(\mbox{Tr} [\tilde{\mathbf W}_v]-\lambda_{\max}(\tilde{\mathbf W}_v)\right)
\\ \notag
&&\quad$Subject to:$  
\notag\\&&\quad\quad\mbox{Tr}\left[\tilde{
\mathbf W}_{v}^T \mathbf B_{v,m}\right]
\ge 1
+\ell_v\left(\tilde{\mathbf g}_{m} 
\right) 
,\quad 1\le m\le M
\notag\\ \notag 
&& \quad\quad\tilde{\mathbf W}_v\ge 0 
, (\tilde{
\mathbf W}_{v})_{j,j}=1, 
\quad 1\le j \le 1+\tilde N_{v-1}. 
\end{IEEEeqnarray}
Theorem 1 in \cite{good_phan2012beamforming} 
guarantees that for large enough $\mu$ Problems \eqref{e:B_const} and \eqref{e:mu_const} are equivalent. As \eqref{e:mu_const} is still not convex, we follow \cite{good_phan2012beamforming} again and solve iteratively using $\mathbf w_{\max}^{(u)} \mathbf w_{\max}^{(u)H}$ as a sub-gradient of $\lambda_{\max}(\tilde{\mathbf W}_v)$ where $\mathbf w_{\max}^{(u)}$ is the unit-norm vector that corresponds to the maximal eigenvalue of $\tilde{\mathbf W}_v$ in iteration $u$. Thus the resulting optimization problem for iteration $u+1$ becomes:
\begin{IEEEeqnarray}{rCl}\label{e:new_mu_const}
&&\tilde{\mathbf W}_v^{(u+1)}=\argmin_{\tilde{
\mathbf W}_v} \, \max_j \, \mbox{Tr}\big[\tilde{
\mathbf W}_{v}^T \mathbf A_{v,j}\big]+\ell_v\left(\mathbf{e}_{j+1}\right)
\notag\\\notag&&\hspace{1cm}+\mu \big(\mbox{Tr} [\tilde{\mathbf W}_v]-\mbox{Tr}(\mathbf w_{\max}^{(u)} \mathbf w_{\max}^{(u)H}(\tilde{\mathbf W}_v-\tilde{\mathbf W}_v^{(u)}))\big)
\notag\\ 
&&\quad$Subject to:$  
\\&&\quad\quad\mbox{Tr}\left[\tilde{
\mathbf W}_{v}^T \mathbf B_{v,m}\right]
\ge 1
+\ell_v\left(\tilde{\mathbf g}_{m} 
\right) 
,\quad 1\le m\le M
\notag\\ \notag 
&& \quad\quad\tilde{\mathbf W}_v\ge 0 
, (\tilde{
\mathbf W}_{v})_{j,j}=1, 
\quad 1\le j \le 1+\tilde N_{v-1}. 
\end{IEEEeqnarray}
and 
the solution of 
\eqref{e:min_power} can be obtained by Algorithm \ref{alg:2}.

\begin{algorithm}
\caption{Evaluating $P_v(\underline{\mathbf w}_v,\eta)$}
\label{alg:2}
\begin{algorithmic}[1]

\State Set $\tilde{\mathbf W}_v^{(0)}$ based on the last last known ${\mathbf w}_v$.
\State Set $\mu=0.25$.
\Repeat
\State Set $\mu=2\mu$, $u=0$.
\Repeat 
\State Set $\mathbf w_{\max}^{(u)}$ as a unit-norm vector correspond-
\State \quad ing to the maximal eigenvalue of $\tilde{\mathbf W}_v^{(u)}$ .
\State Solve \eqref{e:new_mu_const}.
\State set $u=u+1$.
\Until Convergence
\State Set $\tilde{\mathbf W}_v=\tilde{\mathbf W}_v^{(0)}=\tilde{\mathbf W}_v^{(u)}$.
\Until $\mbox{rank} (\tilde{\mathbf W}_v)=1$.
\end{algorithmic}
\end{algorithm}




\section{Deep-learning optimization}\label{sec:DR_optimization}


\subsection{Optimization approach} In our novel approach, we use deep-learning training for the optimization of the network. This training allows the relays to use their non-linear regime, as long as the total network performance increases. 

We should note that there is a conceptual difference between the training of the relay network as opposed to neural networks. In the relay case, the network input contains very little information about the desired functionality. For example, in table \ref{table:modulation}, the network input has only $4$ possible values. As the inputs propagate through
the network they accumulate noise, which is the main adversary in this scenario. Thus, the information on the network is obtained by transmitting each of the few possible inputs many times over the network, and experiencing many different noises and different outputs. 

\subsection{Training}
The network training is based on transmitting a batch of symbols over the network and observing the input and output of each relay and the signals received at the receivers. This resembles the training of a neural network that relies on labeled
data to learn its mapping. 

It is important to note that (unlike in the previous section) the training is done for a specific modulation. Thus, the system input is the unmodulated data bits vector $\bf u$, and the modulation operation is considered part of the system. The system output is the bit estimates, $\hat u_{m,b}$ given in \eqref{e:bit-estimates} and the performance measure is the BER given in \eqref{e:BER1}. However, for training, we take the processed outputs before the final decisions, i.e., $\bar q_{m,b}[k]$ defined in \eqref{e:q_eval_per_bit}.

We denote the network operation as a function of its input and trainable parameters:
\begin{IEEEeqnarray}{rCl}\label{e:XXX}
{\bar{\bf q}}[k]=\mathfrak{f}({\bf u}[k];{\bm\varphi} )
\end{IEEEeqnarray}
where $\bar {\bf q}[k]=[\bar q_{1,1}[k], \ldots, \bar q_{1,B}[k],\bar q_{2,1}[k],\ldots, \bar q_{M,B}[k]]^T$ and the trainable parameters are collected into ${\bm \varphi}=[{\bf w}_1^T,\ldots,{\bf w}_d^T,\bar{\bf w}^T, {\bf b}_1^T,\ldots,{\bf b}_d^T,\bar{\bf b}^T]^T$ with $\bar{\bf w}=[\bar w_1,\ldots \bar w_M]^T$ and $\bar{\bf b}=[\bar b_1,\ldots \bar b_M]^T$. Note that (unlike most neural networks) the function $\mathfrak f()$ is a random function due to the effect of the noise (see \eqref{e:input_first_layer} and \eqref{e:input_general_layer}). The functionality of $\mathfrak f()$ is determined by the network topology as defined by the channel gains $\mathbf h_{i}$, $\mathbf F_{i,\ell}$ and $\mathbf g_{i,m}$, $i=1,\ldots,d$ $\ell=1,\ldots, d-1$ and $m=1,\ldots, M$.

The network optimization can be performed in a distributed online manner, or, in a centralized batch manner. In this work, we focus on the centralized version and leave the distributed version for future research. Yet, it is important to note that the distributed version has two important advantages: i. the backpropagation algorithm has a natural and efficient distributed implementation and ii. online training does not require explicit channel estimations. 

The centralized network training is based on the digital twin paradigm, where the central processor optimizes a simulated version of the network, and then, the optimized parameters are fed into the actual network. Thus, the centralized optimization requires a central processor that knows all channel gains, but, does not require additional pilot transmissions except for those needed for the channel estimation.

We collect the data of $K=600$ (simulated) symbols into a single batch. 
 The training is performed by minimizing the loss function, where we first apply a sigmoid for each output, 
\begin{IEEEeqnarray}{rCl}\label{e:XXX}
\bar u_{m,b}[k] = \sigma(\beta\bar q_{m,b}[k]) = \frac{1}{1+e^{-\beta\cdot\bar q_{m,b}[k]}}
\end{IEEEeqnarray}
with $\beta=5$ 
and then the binary cross entropy loss:
\begin{IEEEeqnarray}{rCl}\label{e:XXX}
\mathcal L_m({\bm \varphi})&=&\frac{1}{BK }\sum_{b=1}^B \sum_{k=1}^K -u_{m,b}[k]\log_2(\bar u_{m,b}[k]) 
\notag\\&&-
(1-u_{m,b}[k])\log_2(1-\bar u_{m,b}[k]).           
\end{IEEEeqnarray}
To combine the BER of the different users, we use the Boltzmann softmax operator:
\begin{IEEEeqnarray}{rCl}\label{e:XXX}
\mathcal L({\bm \varphi}) = \sum_m 
\frac{\mathcal L_m({\bm \varphi}) e^{\alpha \mathcal L_m({\bm \varphi})} }{\sum_me^{\alpha \mathcal L_m({\bm \varphi})}}
\end{IEEEeqnarray}
with $\alpha=5$.

The training module follows the backpropagation algorithm, using gradient-based minimization of the loss with iterations of the form
\begin{IEEEeqnarray}{rCl}\label{e:XXX}
{\bm \varphi}^{(t+1)}={\bm \varphi}^{(t)}-\eta \nabla_{\bm \varphi}\mathcal{L}({\bm \varphi} ^{(t)}).
\end{IEEEeqnarray}
Note that all the training is performed over a single batch of data. Thus, during the training 
the noises ($\mathbf n_i[k] $ and $\tilde{ n}_m
[k] $) are fixed and hence the loss is deterministic. The step size, $\eta$ is updated using the ADAM optimizer \cite{kingma2014adam}.
%

The training optimizes the performance at the specific network setting. To obtain network gains that give good performance over a large range of SNRs, we train the network with adaptive noise variance. Specifically, we start with very low variance, and increase the variance by a factor of $1.5$ whenever the BER goes below $5\%$.

\subsection{Testing and validation}
We do not need to set aside data for testing and validation. These operations are always performed using new random noises (as we can always generate more data).

\subsection{Initialization}
The most direct approach to initialize the training is by using the result of the linear optimization of Section \ref{Sec:Linear}. That is, we use the optimal gains obtained from \eqref{e:linear_opt} and initialize all biases to $\mathbf b_i=\mathbf 0$. But, the solution of \eqref{e:linear_opt} requires a sequential solution of many convex problems and hence is quite complicated for large networks.

Instead, we present here a simpler initialization which showed good performance in our numerical study. The main idea of this initialization is to keep all relays close to their linear regime, so that no relay starts saturated.

We start with layer $i=1$ and then move forward. For relay $n$ of layer $i$ we calculate 
$$p_{i,n}=E[y_{i,n}^2].$$
(Recall that $p_{i,n}$ depends only on the gains and biases of previous layers). Then we draw a random sign $s_{i,n}\in\{-1,1\}$ and random amplitude $ a_{i,n}$ which is uniformly distributed over $[0.5,1] $, and set:
\begin{IEEEeqnarray}{rCl}\label{e:XXX}
w_{i,n}=\frac{a_{i,n}s_{i,n}}{p_{i,n}},\quad b_{i,n}=0.
\end{IEEEeqnarray}
Going over all layers from $i=1$ till $i=d$, we establish a starting point with good dynamic behavior.
 
\begin{figure}[t]
\centerline{\epsfig{figure=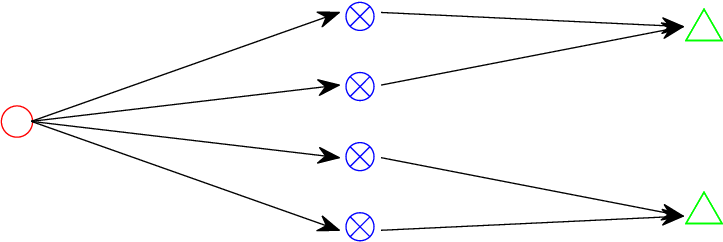,width=.47\textwidth}}
\caption{\label{fig:1layer} A relay network with 4 relays in a single layer.}
\end{figure}
\begin{figure}[t]
\centerline{\epsfig{figure=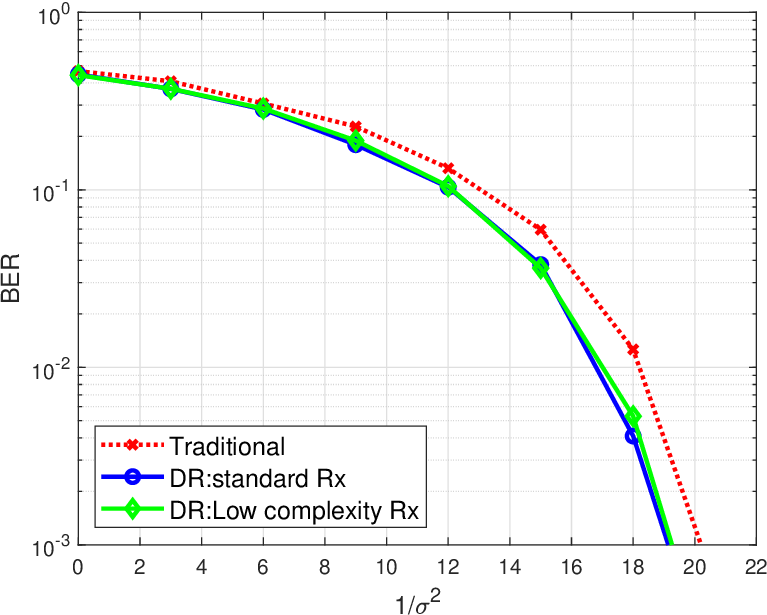,width=.45\textwidth}}
\caption{\label{fig:BER_1_layer} BER vs. $1/\sigma^2$ for the relay network of Fig. \ref{fig:1layer}. The figure compares traditional (linear) optimization and deep-learning optimization (DR).}
\vspace{-3mm}
\end{figure}

\section{Numerical results}\label{sec:numerical}

\subsection{Simple networks}
In this section, we demonstrate the advantages of the use of many relays, and in particular when using our novel optimization approach. We start by studying simple networks with one and two layers. 

As a first example, we consider the network depicted in Fig. \ref{fig:1layer}, with $N=4$ relays and $M=2$ outputs, where each output sees different two relays. In this network, we set all physical channel gains to $1$, that is $\mathbf{h}_1=[1,1,1,1]^T,$ $\mathbf g_{1,1}=[1,1,0,0]^T$ and $\mathbf g_{1,2}=[0,0,1,1]^T$. The resulting BER is depicted in Fig. \ref{fig:BER_1_layer} as a function of $1/\sigma^2$.

\begin{figure}[t]
\centerline{\epsfig{figure=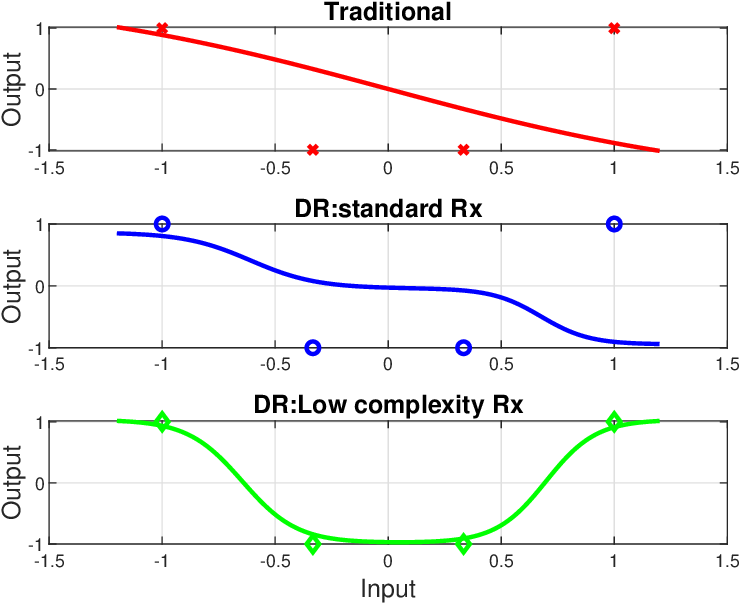,width=.45\textwidth}}
\caption{\label{fig:transfer_1_layers} Network transfer function for user 2: value received by user 2 as a function of the network input in the absence of noise for two optimization approaches and different receivers.}
\vspace{-3mm}
\end{figure}

Fig. \ref{fig:BER_1_layer} shows the BER of the worst user (out of 2) in $3$ scenarios. The red curve (with $x$-marks) depicts the performance with the novel optimization scheme of section \ref{Sec:Linear}, using the traditional linearization approach (i.e., where the optimization treats the relays as linear amplifiers with a power constraint, and we set $P_{\max}=0.64$). The blue and green curves show the results of a deep-learning relay optimization (DR). The blue curve (with circles) shows the performance with standard receivers, i.e., where user $2$ employs a 4 PAM receiver according to Table I. The green curve (with diamonds) shows the performance with low-complexity receivers, i.e., where both users employ a BPSK receiver. 

The figure shows that the deep-learning optimization approach outperforms traditional optimization by 1dB. It is important to highlight the dual significance of this achievement. Firstly, it demonstrates the superiority of non-linear optimization over traditional linear-based optimization techniques. Secondly, it underscores the ability of our approach to deliver enhanced performance while accommodating low-complexity receivers. Specifically, our network enables User 2 to receive a simple BPSK modulation while effectively mitigating the impact of User 1's data on User 2's reception. This result demonstrates the capability of our approach to enhance performance while accommodating simplified receiver configurations even in such a shallow network.

To gain deeper insights into the advantages of non-linear optimization, Fig. \ref{fig:transfer_1_layers} depicts the transfer function of User 2 in each of the optimized networks. The transfer function is the relationship between the value measured at the receiver of User 2 and the input value at the transmitter in the absence of noise\footnote{It is important to acknowledge that the transfer function does not provide a complete description of the network, as the introduction of noise occurs at six different points within the network. This observation holds particularly true when the relays operate in their non-linear regime, where the influence of noise cannot be approximated by a single noise gain.}. The markers in the figure correspond to the transmitted values (on the $x$-axix) and the desired bit values of User 2, where $+1$ represents a 0-bit and $-1$ represents a 1-bit.

\begin{figure}[t]
\centerline{\epsfig{figure=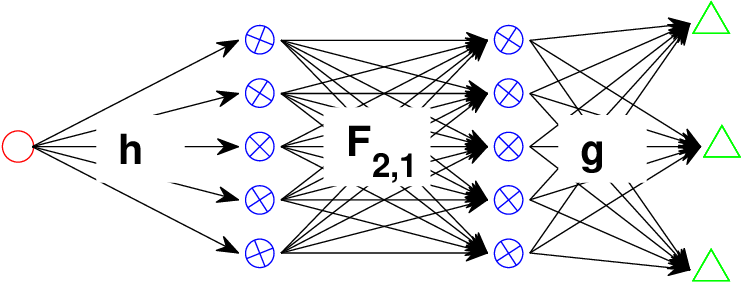,width=.47\textwidth,trim=0 0 0cm 0cm, clip}}
\caption{\label{fig:2layers} A relay network with 2 layers each with 5 relays.}
\end{figure}
\begin{figure}[t]
\centerline{\epsfig{figure=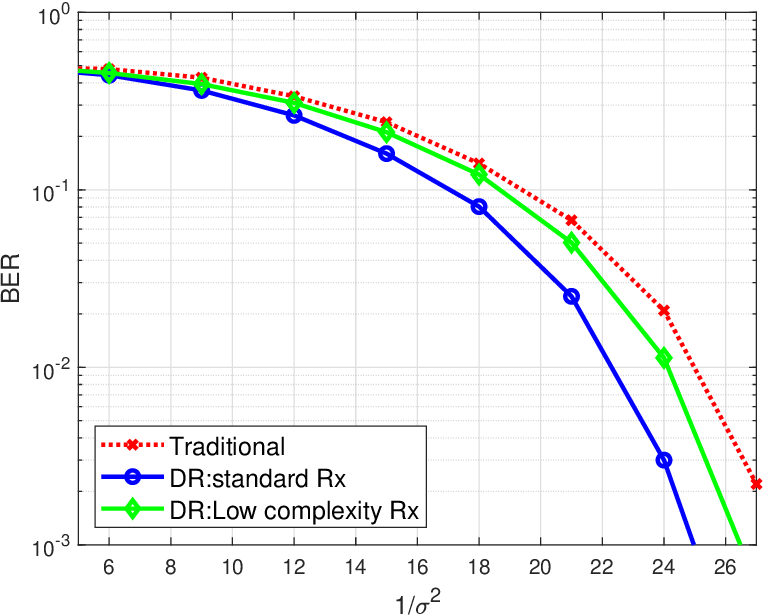,width=.45\textwidth}}
\caption{\label{fig:transfer_2_layers} BER vs. $1/\sigma^2$ for the relay network of Fig. \ref{fig:2layers}. The figure compares traditional (linear) optimization and deep-learning relay (DR) optimization.}
\vspace{-3mm}
\end{figure}
\begin{figure}[t]
\centerline{\epsfig{figure=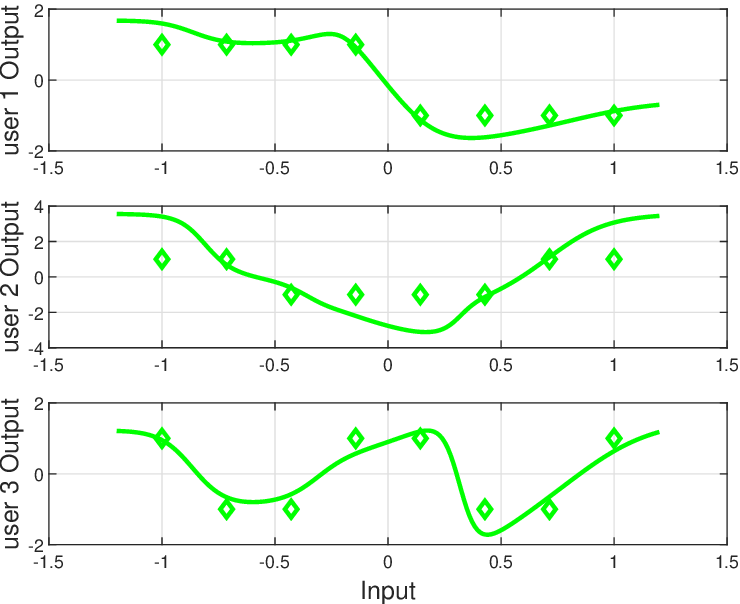,width=.45\textwidth}}
\caption{\label{fig:transfer_2_layers} Network transfer function low-complexity receivers: The value received by each user using the deep-learning relay (DR) optimization with low-complexity receivers.}
\vspace{-3mm}
\end{figure}

The figure shows that the traditional optimization indeed results in a nearly linear transfer function. Recall that the output processing function for this user, $f^{(1)}(-3 \bar r_{2}[k]/4)$ maps values between $-2/3$ and $2/3$ to negative outputs, and hence (with small enough noise) the receiver can detect the transmitted bit.

With the same receiver types, the DR optimization achieves its improvement by merging the two middle modulation points. As a result, both $-1/3$ and $+1/3$ are mapped to values close to zero, while the $-1$ and $+1$ points are mapped to values close to $-1$ and $+1$. This particular mapping improves the detection of the information of User 2, by distinguishing whether the transmitted value originated from one of the intermediate points or from one of the extreme points.

For the utilization of a low-complexity receiver, it is imperative for User 2's data to be effectively separated from the data of User 1. The bottom subplot in Fig. \ref{fig:transfer_1_layers} demonstrates that this separation is indeed accomplished. We should emphasize that once such a separation is achieved, the amplification of the resulting bi-podal signal becomes significantly more effective. Consequently, networks incorporating more layers with DR optimization are anticipated to achieve substantially greater gains compared to those relying on traditional optimization methods.

We proceed to analyze a (slightly) deeper network comprising two layers and three receivers, as depicted in Fig. \ref{fig:2layers}. For this setup, we use: ${\bf h}_1=[1,1,1,1,1]^T$, ${\bf h}_2={\bf 0}$, $\mathbf g_{2,1}=[4,-1,0,0,1]^T$, $\mathbf g_{2,2}=[0,1,4,-1,0]^T$, $\mathbf g_{2,3}=[-1,0,0,1,4]^T$, $\mathbf g_{1,1}=\mathbf g_{1,2}=\mathbf g_{1,3}=\mathbf{0}$, and:
\begin{IEEEeqnarray}{rCl}\label{e:XXX}
\mathbf{F}_{2,1}=\begin{bmatrix}
 1  & -0.5 &  -1 &  -0.5 &   1 \\
   -0.5 &   1 &  -0.5 &   1 &  -0.5\\
   -1 &  -0.5 &   1 &  -0.5 &  -1\\
   -0.5 &   1 &  -0.5 &   1 &  -0.5\\
    1 &  -0.5 &  -1 &  -0.5 &   1
\end{bmatrix}.
\end{IEEEeqnarray}

\begin{figure}[t]
\centerline{\epsfig{figure=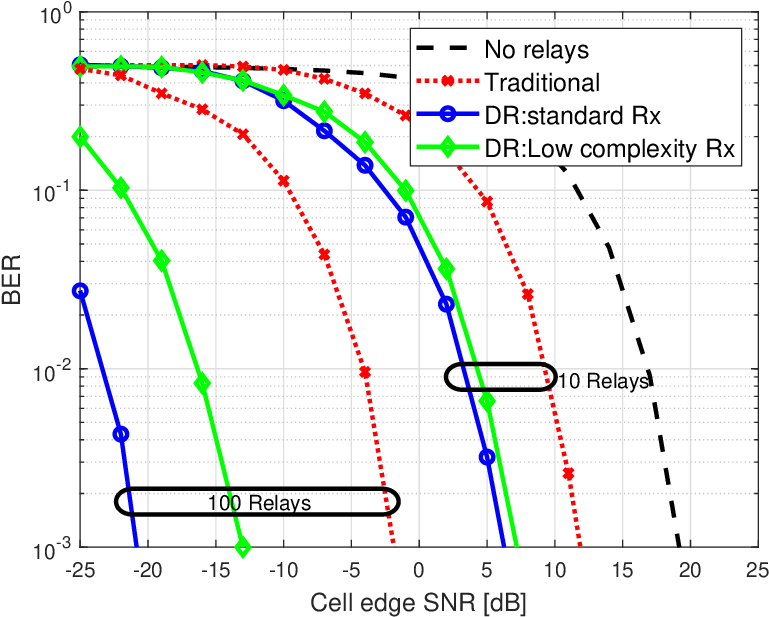,width=.45\textwidth}}
\caption{\label{fig:BER_10_100} BER vs. cell edge SNR for networks with uniformly distributed relays. For reference, the figure also shows the BER curve with no relays in the network.}
\vspace{-3mm}
\end{figure}

The resulting minimal bit error rate (BER) across the three users is depicted in Figure \ref{fig:transfer_2_layers}. Notably, our innovative DR optimization yields a significant gain of approximately 3dB over the traditional method when employing standard receivers. This gain is noteworthy, as it enables us to reduce all transmission powers by a factor of two at the only price of a more intelligent optimization technique.

In this scenario, achieving effective data separation for three users poses a more intricate challenge. Consequently, the network that utilizes low-complexity receivers exhibits slightly inferior performance compared to the one employing standard receivers. Nevertheless, even with low-complexity receivers, DR optimization achieves a gain of around 1dB over traditional optimization (where the latter utilizes standard PAM receivers). 

Fig. \ref{fig:transfer_2_layers} demonstrates the successful data separation for all three users in the low-complexity DR network. The figure depicts the transfer functions of the three receivers and shows that the network indeed implemented the required receiver structure for each user (the markers denote the desired output sign for each receiver).

The apparent asymmetry observed in the curves in Fig. \ref{fig:transfer_2_layers} may raise concerns regarding the effectiveness of the network. However, we can abate such concerns by recalling that the receivers only take the sign of the received signals, and that all zero crossings in the curves of Fig. \ref{fig:transfer_2_layers} are in close proximity to their optimal positions. Furthermore, we recall that the BER curves in Fig. \ref{fig:transfer_2_layers} exhibit excellent performance, clearly indicating the efficiency of the network. Consequently, we conclude that the observed asymmetry stems from the necessity to implement three distinct functions within the same 10-relay network. 

In this context, it is important to remember that the relay network has lower computational capability than a neural network of the same size. This is because most of the connections in the relay network are determined by the physical conditions and are not trainable. For example, in the network of Fig. \ref{fig:2layers} there are $39$ connections but only $20$ trainable parameters (a gain and a bias for each of the $10$ relays).

\begin{figure}[t]
\centerline{\epsfig{figure=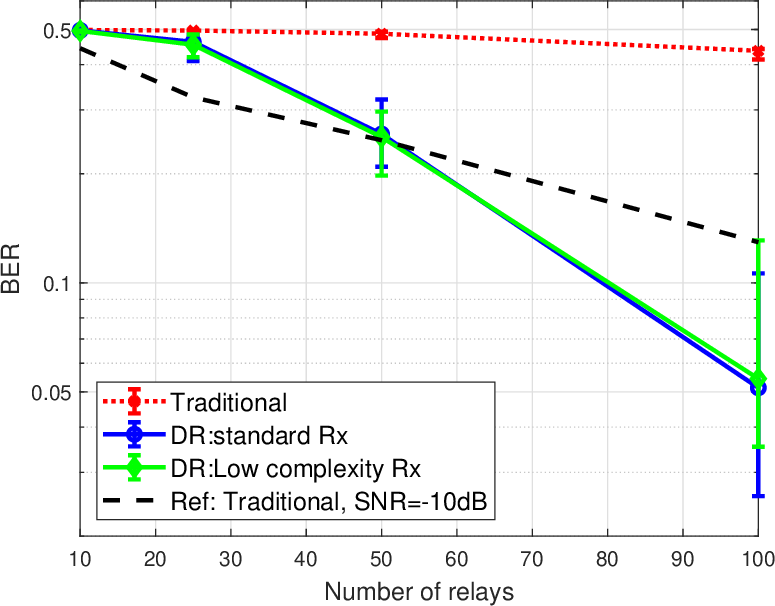,width=.45\textwidth}}
\caption{\label{fig:BER_median} BER vs. the number of relays for networks with uniformly distributed relays at a cell edge SNR of $-22$dB. The lines show the medial BER over 10 random networks. The error bars show the $25$-th and $75$-th percentiles. For reference, the figure also shows the median BER with traditional relay optimization at cell edge SNR of $-10$dB.}
\vspace{-3mm}
\end{figure}

\subsection{Spatial distribution of relays}
We now turn to a more practical model in which the relays are distributed across a two-dimensional plane. The network consists of a single-antenna base station (BS) transmitter, two single-antenna receivers, and 100 relays equipped with directional antennas, as illustrated in Fig. \ref{fig:prelim_system_model}. The network represents a sector spanning $60^\circ$ of a cell with a radius of 100 meters. The receivers are positioned at the cell edge.

To ensure the absence of loops within the network, we assume that the relays use ideal directional antennas.
(in Fig. \ref{fig:prelim_system_model}, the sectors in each relay illustrate the directional beams of their respective antennas.)

We adopt the cell edge SNR as the reference SNR for this model. Using our previous normalization of the transmission power to $1$, the cell edge SNR is given by $10^{-8}/\sigma^2$.

 The resulting BER curves (of the worst out of the 2 users) are depicted in Fig. \ref{fig:BER_10_100} for a network of 10 relays and a network of 100 relays. For reference, the figure also shows the BER curve in a network with no relays. The figure clearly shows the gain from the use of many relays. Considering, for example, a BER of $0.01$, we see that with traditional optimization, the network with $10$ relays gained $8$dB while the network with $100$ relays gained $21$dB over the reference network (without relays). Recalling that the deployment and utilization are fairly easy, such gains show the potential for a huge boost in communication performance. 

More surprisingly, DR optimization shows a huge gain over traditional optimization in large networks. In the network with $100$ relays, the use of DR optimization gained $19$dB over the best-known traditional optimization. This huge gain comes only at the price of recognizing the non-linearity in the relays and using the appropriate deep-learning tools. Thus, we conclude that the DR approach has a huge potential for boosting communication performance in large networks. 

Note that Fig. \ref{fig:BER_10_100} only shows one realization of the random networks. To better reflect the statistical nature of these networks, Fig. \ref{fig:BER_median} shows the median of the BER over $10$ realizations of each network, as a function of the number of relays in the network. All BERs in this figure (except for the reference curve) are evaluated when the cell edge SNR is $-22$dB. The error bars in the figure also show the BER of the first and last quartiles. 

The figure shows that the gain of DR over traditional optimization is negligible for the $10$-relays network, but grows very fast with the number of relays. To better understand this gain, the figure also shows the median BER with traditional optimization, but at a higher SNR of $-10$dB. Thus, we can see that with $50$ relays, the median gain of DR over traditional optimization is larger close to $12$dB. This confirms that the huge gains of Fig. \ref{fig:BER_10_100} are not unique for the specific realization. Moreover, with $100$ relays, the vast majority of network realizations achieved a gain of more than $12$ dB over traditional optimization.

\section{Conclusions}
We presented a novel approach for the optimization of relay networks. Unlike the traditional approach that approximates relays as linear amplifiers, our novel approach takes into account the true non-linear nature of the relays. Using the similarity between the transfer function of a relay and the transfer function of a neuron, we employ deep-learning methodology to better optimize the network. Numerical study shows huge gains compared to traditional optimization.

This paper focused on the optimization of cascade relay networks. We first solved the optimization in the traditional approach, i.e., treating relays as linear amplifiers. In this approach, we formulated the min-max BER optimization problem and presented a novel algorithm with guaranteed convergence, at least to a local maximum.

Then, we introduced the novel DR optimization approach for the optimization of relay networks. Departing from the `linear amplifiers' paradigm and leveraging deep-learning techniques we introduced a completely new approach for relay network optimization. Numerical studies demonstrated significant performance gains compared to traditional optimization methods. For large networks, these gains were shown to often exceed $10$dB.

Moreover, we explored the capability of non-linear relay networks to implement versatile functions, paving the way for the implementation of new functionalities over the network. As an example, we demonstrated the non-linear separation of data for different users, which reduced the receiver complexity and improved the data delivery.

As a pioneering work on deep relay optimization, our primary objective was to unveil the potential of this approach and reveal its fundamental characteristics.
Hence, this work took simplifying assumptions that allowed us to better focus on the core issues of this network. 

Future research is required to address practical concerns such as proper learning with complex signals and gains, 
imperfect directional antennas and gain loops.
Also, additional research is required to determine the effect of the network structure on learning flexibility (recalling that, unlike neural networks, here the network structure is determined by the physical channel gains between relays).


\bibliographystyle{ieeetr}
\bibliography{DR_downlink}

\end{document}